\documentclass[10pt,a4paper]{article}
\usepackage{amsmath,latexsym,amssymb,amsfonts}
\usepackage[dvips]{graphicx}
\usepackage{bm}

\begin{document}

\title{\textbf{The no-boundary measure in string theory}\\ \large{\textsf{Applications to moduli stabilization, flux compactification, and cosmic landscape}}}
\author{\textsc{Dong-il Hwang}$^{a,b}$,\;\; \textsc{Bum-Hoon Lee}$^{b}$,\;\; \textsc{Hanno Sahlmann}$^{c,d,e}$\;\;\\
and\;\; \textsc{Dong-han Yeom}$^{b,f}$\\
\textit{$^{a}$\small{Department of Physics, KAIST, Daejeon 305-701, Republic of Korea}}\\
\textit{$^{b}$\small{Center for Quantum Spacetime, Sogang University, Seoul 121-742, Republic of Korea}}\\
\textit{$^{c}$\small{Asia Pacific Center for Theoretical Physics, Pohang 790-784, Republic of Korea}}\\
\textit{$^{d}$\small{Physics Department, Pohang University for Science and Technology, Pohang 790-784, Republic of Korea}}\\
\textit{$^{e}$\small{Department of Physics, University Erlangen-Nueremberg, 91058 Erlangen, Germany}}\\
\textit{$^{f}$\small{Research Institute for Basic Science, Sogang University, Seoul 121-742, Republic of Korea}}
}
\maketitle

\begin{abstract}
We investigate the no-boundary measure in the context of moduli stabilization. To this end, we first show that for exponential potentials, there are no classical histories once the slope exceeds a critical value. We also investigate the probability distributions given by the no-boundary wave function near maxima of the potential. These results are then applied to a simple model that compactifies $6$D to $4$D (HBSV model) with fluxes.
We find that the no-boundary wave function effectively stabilizes the moduli of the model.

Moreover, we find the \textit{a priori} probability for the cosmological constant in this model. We find that a negative value is preferred, and a vanishing cosmological constant is not distinguished by the probability measure.

We also discuss the application to the cosmic landscape. Our preliminary arguments indicate that the probability of obtaining anti de Sitter space is vastly greater than for de Sitter.
\end{abstract}

\newpage
\tableofcontents
\newpage

\section{Introduction}

String theory predicts the existence of extra dimensions beyond the observed four. To conform with observations, these must be compactified to very small scales. The geometry of the extra dimensions is dynamical in principle, thus the dynamics has to explain why the extra dimensions stay curled up.
In low energy effective actions, the geometry it is described by scalar fields evolving in complicated potentials, and the problem of trapping them is known as the moduli stabilization problem. In the context of quantum theory, the question is whether the fields find themselves in a state that is sufficiently and suitably localized in positions and momenta to be trapped in a local minimum of the potential.
Here the initial state comes into play, because it determines probabilities for such dynamical conditions to be satisfied. In the present work, we analyze the ability of a specific proposal for the initial state, the no boundary wave function of Hartle and Hawking in this respect. To be precise, we are calculating approximate probabilities for the field to find itself as a sharply peaked wave packet at various locations in the potential. These can then be used for further dynamical considerations because such a wave packet then evolves further according to the classical equations of motion and may get trapped by a local minimum, or roll in an unstable direction forever. In the following, we describe the moduli stabilization problem and the no boundary proposal in more detail.

\subsection{Flux compactification in string theory}

\paragraph{Moduli stabilization problem}

String theory is motivated to resolve the renormalization problem \cite{Polchinski:1998rq}. Einstein gravity is not renormalizable; while, if we introduce string theory, then all quantum corrections can be manifestly finite. However, to obtain a consistent quantum theory, (super) string theory requires ten dimensions.

A rather traditional approach to reducing dimensions is to compactify extra dimensions. It is known that a Calabi-Yau manifold can maintain supersymmetry, so that it can be possible to obtain a four-dimensional effective supersymmetric action. Another approach is not to compactify extra dimensions, and to stipulate that we are attached to a D-brane; extra dimensions can be sufficiently large, or extra dimensions can be warped so that all gauge fields can be confined on three-branes. The latter approach is a viable and interesting idea, but the existence of such a brane combination and a metric ansatz should be explained in the context of string theory.

In terms of a bottom-up approach, the traditional compactifying scenario is rather rigorous. However, there is a well known problem, that of \textit{moduli stabilization}. When we compactify extra dimensions to a compact manifold, there are continuous degrees of freedom that determine the size or shape of the compact manifold. After integrating out the extra dimensions, the degrees of freedom become a number of fields in the four-dimensional effective action, the moduli fields. There is no fundamental mechanism to stabilize these fields, and thus no fundamental mechanism keep the size and shape of the compactified manifolds fixed. Of course, one may believe that the fields can be initially fixed to have a certain value. However, there is no way to prohibit the movement of the fields during an evolution of the universe, and then it becomes very strange to see a stable compactified universe at the present time.

A solution of this problem requires at least two conditions to be satisfied:
\begin{enumerate}
\item All moduli fields must have potentials and the potentials must have local minima.
\item If the potential has a non-compact region in which the sign of the slope is negative (`run-away region') the probability for the system to end up in this region should be strictly smaller than one.
\end{enumerate}
The best-known mechanism for satisfying the first condition is the so-called \textit{flux compactification} \cite{Bousso:2000xa}, which we will discuss in the next paragraph. In this article, for a given toy model of flux compactification, we will try to determine whether the second condition is met for a universe born from the no-boundary wave function.

\paragraph{Flux compactification}

To induce fluxes, we require some charged objects in the manifold. Good candidates for this purpose are D-branes, physical objects with tension, mass, and charge.

In general, the natural tendency of moduli fields is to collapse so that they tend to become smaller and smaller. However, the existence of flux tends to enlarge the size of the manifold. So, there can be a stable equilibrium, in which two forces -- collapsing and repulsing -- cancel each other.

This flux compactification is an important mechanism to stabilize moduli fields. However, the properties of the induced potential now become important, in particular the potential energies at the local minima. Anti de Sitter vacua are quite natural in string theory, but our universe is not described by anti de Sitter space. To obtain a de Sitter vacuum, one has to violate supersymmetry.

One successive model to violate supersymmetry and to obtain a de Sitter vacuum is the model of Kachru, Kallosh, Linde, and Trivedi \cite{Kachru:2003aw}. In their model, they introduced D3-branes and anti D3-branes. Since they require tadpole cancelation, the only quantum effects in their calculation are non-perturbative. They violate supersymmetry and eventually lift up the potential from anti de Sitter to de Sitter.

The level of uplift is determined by the number of anti D3-branes. The number of anti D3-branes is a free parameter, and hence each value of the vacuum can be chosen arbitrarily. As the number of fluxes changes, so does the allowed vacuum energy. The number of vacua can be huge, so that there is effectively a continuous spectrum of cosmological constants (so called \emph{discrituum} \cite{Susskind:2003kw}).

In this article, we study a simple model (Halliwell \cite{Halliwell:1985bx} and Blanco-Pillado, Schwartz-Perlov, and Vilenkin \cite{BlancoPillado:2009di}), which we will call the HBSV model. Let us begin with the six-dimensional action:
\begin{eqnarray}
S = \frac{1}{16\pi G_{6}} \int d^{6}x \sqrt{-g} \left( \mathcal{R}^{(6)} - 2\Lambda^{(6)} \right) - \frac{1}{4} \int d^{6}x \sqrt{-g} F_{MN}F^{MN},
\end{eqnarray}
where $M,N=0,1,...,5$ are coordinates, $\Lambda^{(6)}$ is the six-dimensional cosmological constant, and $F_{MN}$ is the field strength tensor of the Maxwell field. Now, assuming the metric ansatz
\begin{eqnarray}
ds^{2} = g_{MN}dx^{M}dx^{N} = e^{-(\sqrt{2G_{6}}/R)\psi(x)} g_{\mu\nu}dx^{\mu}dx^{\nu} + e^{(\sqrt{2G_{6}}/R)\psi(x)}R^{2} d\Omega_{2}^{2}
\end{eqnarray}
and the Maxwell field
\begin{eqnarray}
A_{\tilde{\varphi}} = -\frac{n}{2e} \left(\cos\tilde{\theta} \pm 1\right),
\end{eqnarray}
where $\mu,\nu = 0, 1, 2, 3$ with $[t,\chi,\theta,\varphi]$ and the compactified coordinates are just a sphere $[\tilde{\theta},\tilde{\varphi}]$, we obtain four-dimensional effective action after integration out:
\begin{eqnarray}
S = \int d^{4}x \sqrt{-g} \left( \frac{1}{16\pi G_{4}} \mathcal{R}^{(4)} - \frac{1}{2} (\nabla \psi)^{2} - V_{n}(\psi) \right)
\end{eqnarray}
and
\begin{eqnarray}
V_{n}(\psi) = \frac{1}{2} \left( \frac{\pi n^{2}}{e^{2} R^{2}} e^{-3(\sqrt{2G_{6}}/R)\psi} - \frac{1}{G_{6}}e^{-2(\sqrt{2G_{6}}/R)\psi} + R^{2}\Lambda^{(6)} e^{-(\sqrt{2G_{6}}/R)\psi} \right),
\end{eqnarray}
where $G_{4}=G_{6}/V_{2}$ and $V_{2}=4\pi R^{2}$. In this article, we choose the four-dimensional Planck units $\hbar = c= G_{4}=1$. Then we have
\begin{eqnarray}
V_{n}(\psi) = \frac{\pi n^{2}}{2e^{2} R^{2}} e^{-3\sqrt{8\pi}\psi} - \frac{1}{8\pi R^{2}}e^{-2\sqrt{8\pi} \psi} + \frac{R^{2}\Lambda^{(6)}}{2} e^{- \sqrt{8\pi} \psi}.
\end{eqnarray}
The HBSV model has two essential properties in this context: (1) It has one moduli field that is stabilized by a potential and (2) as one increases the number of fluxes, one can change the vacua from anti de Sitter to de Sitter. The former property is useful to explain the moduli stabilization problem and the latter is useful to understand the basic nature of huge numbers of flux vacua, the so-called cosmic landscape.

\paragraph{Cosmic landscape and multiverse}

As one changes the number of fluxes, one can see different vacuum expectation values. According to Susskind \cite{Susskind:2003kw}, each different flux can be approximated by some fields, so that each vacuum corresponds to different field values according to the complex potential of the fields. Then, tunneling from one vacuum to another vacuum can be described by the tunneling of a scalar field with a certain potential. The huge number of different vacua that can be approximated by the potential of a scalar field is called by the \textit{cosmic landscape}.

The huge number of different vacua can be on the order of $10^{500}$. Then, this can explain the fine-tuning problem of the cosmological constant, in that it needs fine-tuning on the order of $10^{-120}$. Therefore, the cosmic landscape can explain that there can be a huge number of different universes. However, this does not necessarily imply that such universes should exist. To realize a vacuum of the landscape, Susskind considered the eternal inflation. Since an eternal inflation would never end, at a certain time, tunneling will realize \textit{all} possible vacua, as pocket universes. The totality of pocket universes are called the \textit{multiverse}, and, by definition, the multiverse has infinite volume and is formed by an infinite number of all possible pocket universes.

If the multiverse picture becomes a concrete scientific hypothesis, it can assign probabilities to each pocket universe \cite{Linde:2006nw}. However, it is known that such a measure is difficult to define \cite{Aguirre:2006ak}. There are some promising candidates; however, there is no common consensus on the problem.

The wave function of the universe has some interesting positions in this context, as it  \cite{Hartle:1983ai} encodes the probability of a universe with certain initial conditions (also, as historic references: \cite{Hawking:1984hk}). Will it prefer eternal inflation? Will it form the multiverse? Can it have implications for the measure problem of the multiverse? At least, Hartle, Hawking, and Hertog \cite{Hartle:2007gi}\cite{Hartle:2008ng}\cite{Hartle:2010vi}\cite{Hartle:2010dq} believed so, and they argued that the no-boundary wave function does not prefer a multiverse \cite{Hartle:2007gi}\cite{Hartle:2008ng}, and even though there is eternal inflation, the no-boundary measure is still meaningful and it will give some expectations about the multiverse \cite{Hartle:2010vi}\cite{Hartle:2010dq}. On the other hand, some authors have believed that the expectation of the no-boundary proposal is contradictory with the multiverse picture \cite{Dyson:2002pf}\cite{Page:2006hr} and hence the wave function of the universe is not useful to understand the nature of the multiverse measure. We hope that we can critically appraise and answer these opinions.

\paragraph{Purpose of this paper}
Wit the present article, we would like to shed some light on the question of whether the no-boundary proposal is viable in the context of string theory, or not. In the context of the moduli stabilization problem, we commented that two conditions need to be met.
In addition, since we live in a de Sitter universe, the further requirements are natural:
\begin{itemize}
\item[3.] Various vacuum expectation values, including positive vacuum energy, should be allowed, and hence lead to a cosmic landscape.
\item[4.] With a view to observations, a small positive cosmological constant and a sufficent inflationary history should be preferred.
\end{itemize}
The HBSV model satisfies conditions $1$ and $3$. In the current article, we will ask whether the no-boundary wave function can explain conditions $2$ and $4$.

To summarize, we will ask, and partially answer, the following questions about the no-boundary wave function and apply it to the following questions:
\begin{description}
\item[Moduli stabilization problem:] Does the no-boundary measure explain the stability of the moduli fields?
\item[\textit{A priori} probability of the cosmological constant:] Does the no-boundary measure explain the \textit{a priori} probability of the cosmological constant, or the probability for each vacuum expectation value?
\item[Probability of the cosmic landscape:] What is the relation between the wave function of the universe and the multiverse measure? Does the no-boundary measure prefer eternal inflation and multiverse? If not, which universe is preferred in the landscape?
\end{description}
To be concrete, we will do calculations within the HBSV model.

\subsection{Review of no-boundary measure}

\paragraph{No-boundary proposal}

The no-boundary wave function \cite{Hartle:1983ai} for gravity coupled to a matter field is
\begin{eqnarray}\label{eqn:noboundary}
\Psi[h_{\mu\nu}, \chi] = \int_{\partial g = h, \partial \phi = \chi} \mathcal{D}g\mathcal{D}\phi \;\; e^{-S_{\mathrm{E}}[g,\phi]},
\end{eqnarray}
where $h_{\mu\nu}$ and $\chi$ are the boundary values of the Euclidean metric $g_{\mu\nu}$ and the matter field $\phi$ which are the integration variables, and the integration is over all non-singular geometries with a single boundary.
$S_{\mathrm{E}}$ is the Euclidean action:
\begin{eqnarray}
S_{\mathrm{E}} = -\int d^{4}x \sqrt{+g} \left( \frac{1}{16\pi}R - \frac{1}{2} (\nabla \phi)^{2} - V(\phi) \right)
\end{eqnarray}
for $\phi$ a scalar field
This path integral formula is a solution of the Wheeler-DeWitt equation. Moreover, can be regarded as a `ground state' for the gravitational field \cite{Hartle:1983ai}.
However, the path integral in (\ref{eqn:noboundary}) badly diverges as the action is not bounded from below, see for example \cite{Vilenkin:1994rn} for a discussion. To obtain convergence, Halliwell and Hartle \cite{Halliwell:1989dy}\cite{Halliwell:1990qr} argued that regarding the path integral as a contour integral and choosing a contour involving \textit{complex} metrics and fields may improve convergence.

In the minisuperspace approximation:
\begin{eqnarray}
ds_{\mathrm{E}}^{2} = d\eta^{2} + \rho^{2}\left(\eta\right) \left(d\chi^{2} + \sin^{2}\chi \left( d\theta^{2} + \sin^{2}\theta d\varphi^{2} \right) \right).
\end{eqnarray}
Then, the minisuperspace version of the no-boundary proposal with complex contour integration is
\begin{eqnarray}
\Psi[a, \chi] = \int_{\mathcal{C}}\mathcal{D}\rho\mathcal{D}\phi \;\; e^{-S_{\mathrm{E}}[\rho,\phi]},
\end{eqnarray}
where
\begin{eqnarray}
S_{\mathrm{E}} = 2 \pi^{2} \int d\eta \left[ -\frac{3}{8\pi} \left( \rho \dot{\rho}^{2} + \rho \right) + \frac{1}{2}\rho^{3} \dot{\phi}^{2} + \rho^{3} V(\phi) \right]
\end{eqnarray}
and the integration is over (potentially complex valued) regular geometries and fields that connect the boundary values
\begin{eqnarray}
\rho|_{\text{boundary}} = a, \;\;\; \phi|_{\text{boundary}}=\chi
\end{eqnarray}
with the `no-boundary' initial conditions
\begin{eqnarray}
\rho|_{\text{initial}} = 0, \;\;\; \dot{\rho}|_{\text{initial}} = 1, \;\;\; \dot{\phi}|_{\text{initial}}=0,
\end{eqnarray}
which express regularity of the geometry at the initial time. The dot in the last equation stands for derivative with respect to any chosen time parameter.

\paragraph{Classicality condition}

Because of the analytic continuation to complex functions, the action is in general complex, so that
\begin{equation}
\Psi[a,\chi] = A[a,\chi] e^{i S[a,\chi]}
%\approx e^{-\mathfrak{Re} S_{\mathrm{E}}/\hbar}e^{-i\mathfrak{Im} S_{\mathrm{E}}/\hbar}.
\label{eq:class}
\end{equation}
with $A,S$ real.
If the rate of change of $S$ is much greater than that of $A$,
\begin{equation} \label{eqn:classicality}
|\nabla_I A(q)|\ll |\nabla_I S(q)|, \qquad I=1,\ldots n,
%\label{eq:}
\end{equation}
%or equivalently,
%\begin{equation}
%|\nabla_I \mathfrak{Re} S_{\mathrm{E}}|\ll |\nabla_I \mathfrak{Im} S_{\mathrm{E}}|, \qquad I=1,\ldots n,
%%\label{eq:}
%\end{equation}
then the wave function describes almost classical behavior \cite{Hartle:2007gi}\cite{Hartle:2008ng}. In fact
the Wigner function $W[\Psi]$ of a state satisfying the \emph{classicality condition} \eqref{eqn:classicality} in some region of $q$-space is approximately
\begin{equation}
W[\Psi](q,p)
\sim |A(q)|^2\,\delta(p-\nabla S).
%\label{eq:}
\end{equation}
This shows that for $q$ in this region, $\Psi$ determines a probability distribution for $q$ \emph{and} a momentum value $p=\nabla S$, which has a high likelihood.

\paragraph{Steepest descent approximation}

To calculate the path integral, we will use the steepest descent approximation, which requires us to evaluate the action for on-shell paths. As usual, to obtain the best approximation, complex saddle points have to be admitted. To solve the equations of motion, we must chose a time parameter, and since we are already forced to consider complex metrics and fields by the steepest descent approximation, it is useful to regard the action integral as a contour integral in complex time, and chose a time parameter that is not always real. We call the on-shell complexified instantons \textit{fuzzy instantons}.

For practical reasons we follow \cite{Hartle:2007gi}\cite{Hartle:2008ng} and consider time contours
that begin parallel to the Euclidean time axis, and at a certain point, turn to the Lorentzian time axis.
We denote Euclidean time with $\eta$, real time with $dt \equiv -id\eta$, and the turn point in Lorentzian time with $X$. At the starting point of the contour, we require the no-boundary initial conditions,
after some time along the Lorentzian direction, we require the classicality condition and that all fields be real.

We solve the classical equations of motion for Euclidean and Lorentzian time directions
\begin{eqnarray}
\label{E3}\ddot{\phi} &=& - 3 \frac{\dot{\rho}}{\rho} \dot{\phi} \pm V',\\
\label{E4}\ddot{\rho} &=& - \frac{8 \pi}{3} \rho \left( \dot{\phi}^{2} \pm V \right),
\end{eqnarray}
where the upper sign is for the Euclidean time and the lower sign is for the Lorentzian time.
The on shell Euclidean action is
\begin{eqnarray}
S_{\,\mathrm{E}} = 4\pi^{2} \int d \eta \left( \rho^{3} V - \frac{3}{8 \pi} \rho \right),
\end{eqnarray}
and after the turning point, we integrate along $d\eta = i dt$.

The required initial conditions are
\begin{eqnarray}
\rho(0)^{\mathfrak{Re}} = \rho(0)^{\mathfrak{Im}} &=& 0,\\
\dot{\rho}(0)^{\mathfrak{Re}} &=& 1,\\
\dot{\rho}(0)^{\mathfrak{Im}} &=& 0,\\
\dot{\phi}(0)^{\mathfrak{Re}} = \dot{\phi}(0)^{\mathfrak{Im}} &=& 0.
\end{eqnarray}
At the junction time $\eta = X$, we paste $\rho(\eta)$ and $\rho(\eta)$ to $\underline{\rho}(t)$ and $\underline{\phi}(t)$ so that
\begin{align}
\label{eq:junction1}\underline{\rho}(t=0) = \rho(\eta=X), \;\;\;\; \underline{\dot{\rho}}(t=0)=i\dot{\rho}(\eta=X),\\
\label{eq:junction2}\underline{\phi}(t=0) = \phi(\eta=X), \;\;\;\; \underline{\dot{\phi}}(t=0)=i\dot{\phi}(\eta=X).
\end{align}
The remaining initial conditions are the initial field value $\phi(0) = \phi_{0} e^{i\theta}$, where $\phi_{0}$ is a positive value and $\theta$ is a phase angle. After fixing $\phi_{0}$, by tuning the two parameters $\theta$ and the turning point $X$, we (1) evolve Equations~(\ref{E3}) and (\ref{E4}), (2) calculate the classicality condition (Equation~(\ref{eqn:classicality})), and (3) find the most optimal initial condition that satisfies the classicality condition\footnote{The detailed numerical technique (searching algorithm) is introduced in \cite{Hwang:2011mp}.}. If the potential is symmetric, there can be multiple solutions; however, after we restrict the final condition (e.g., `$\phi(T)$ is in the left side of the local maximum' for a sufficiently large $T$), then we can uniquely specify the solution. If there exists a classical history, then we can calculate a meaningful probability for a classical universe.

\section{Fuzzy instantons in Einstein gravity}

Now we will investigate some properties of the probability distribution on histories given by the no-boundary wave function for Einstein gravity coupled to a scalar field. This is the preparation for discussion of the resulting probability distributions in potentials of the type that show up in string theory compactifications, and in the HBSV model. For doing calculations we must work in the minisuperspace approximation.

\subsection{Definitions and Notation}

We will denote the space of histories satisfying the no-boundary condition by $\mathfrak{H}$.\footnote{Note that histories which differ by time reparametrization are considered the same. Notice also that while these histories are regular in Euclidean time as per the no-boundary conditions, they may well have singularities along the real time axis.} We eventually want to calculate the probability of histories satisfying certain (boundary-) conditions. Let us say that we are interested in histories satisfying a certain condition $A$. Then we can define the subset by
\begin{equation}
\mathfrak{H}_{A} = \{h \in \mathfrak{H} \;|\; h \mathrm{\, has\, property\,} A\}
\end{equation}
of $\mathfrak{H}$.
Given a set of classical histories $\mathfrak{H}_A$, in principle we have
\begin{equation}
P_{\,A}=\int_{Q_A} |\Psi(h,\chi)|^2 n \cdot \nabla S \, \mathcal{D}\mu(h,\chi).
%\label{eq:}
\end{equation}
The integration is over a subset $Q_A$ of a spatial slice with normal $n$ in superspace.
$Q_A$ is the set of points on this slice such that the wave function satisfies the classicality condition in a way compatible with the condition $A$. $S$ was defined in \eqref{eq:class}. $\mu$ is a certain measure which can be obtained in principle from the inner product on the space of solutions to the Wheeler-DeWitt equation and the slice. But it is very difficult to obtain in practice. Using minisuperspace and steepest descent approximation, using $\Phi_0$ as parameter on the slice, and ignoring details of the measure as well as the variation of  $n \cdot \nabla S$,
\begin{equation}
P_{\,A}\approx \frac{1}{Z_A}\int_{Q_A} |\exp(-S_\text{E}[h_{\phi_0}])|^2 d\phi_0=
\frac{1}{Z_A}\int_{Q_A}e^{-2\mathfrak{Re} S_\text{E}[h_{\phi_0}]} d\phi_0,
\label{eq:prob}
\end{equation}
where $Z_A$ is some normalization constant and $h_{\phi_0}$ is a history that initially has scalar field modulus equal to $\phi_0$.

If we have two conditions $A$, $B$, where $A$ is an initial and $B$ a final condition, we will use the notation $\mathfrak{H}_{A\rightarrow B}$ for $\mathfrak{H}_A\cap \mathfrak{H}_B$, and $P_{\,A\rightarrow B}$ for the corresponding probability.

\subsection{Static results}

Even under minisuperspace and stationary phase approximation, analytic calculations are quite difficult since they involve solving the equations of motion and calculating the action for various initial conditions.
For the quadratic potential $V(\phi)=(1/2)m^{2}\phi^{2}$, an approximate calculation is due to Lyons \cite{Lyons:1992ua}, and since it is instructive, we will discuss it briefly, here.
The starting point are the following approximate solutions of the equations of motion.
\begin{eqnarray}
\phi \simeq \phi_{0} + i \frac{m}{3} \sqrt{\frac{3}{4\pi}}\eta, \;\;\; \rho \simeq \sqrt{\frac{3}{4\pi}}\frac{i}{m\phi^{\mathfrak{Re}}_{0}} \exp \left(-i\sqrt{\frac{4\pi}{3}}m\phi_{0}\eta + \frac{1}{6}m^{2}\eta^{2} \right),
\end{eqnarray}
in which the scalar field $\phi$ slowly rolls.
If the scalar field rolls more slowly, then we can further approximate
\begin{eqnarray}
\label{approx2}\phi \simeq \phi^{\mathfrak{Re}}_{0}, \;\;\; \rho \simeq \sqrt{\frac{3}{4\pi}}\frac{1}{m\phi^{\mathfrak{Re}}_{0}} \sin \left(\sqrt{\frac{4\pi}{3}}m\phi^{\mathfrak{Re}}_{0}\eta \right).
\end{eqnarray}

We choose the integration contour in two steps. $(1)$ We integrate in Euclidean time direction from $\eta^{\mathfrak{Re}} = 0$ to $\eta^{\mathfrak{Re}} = \sqrt{3\pi}/4m\phi^{\mathfrak{Re}}_{0} \equiv X$ so that the imaginary part of $\phi$ vanishes. $(2)$ At the turning point $\eta^{\mathfrak{Re}} = X$, we turn to the Lorentzian time direction.

Using this contour of integration, the Euclidean action can be calculated. Note that, if the classicality condition is valid, the real part of the action $S_{\mathrm{E}}$ picks up the biggest contribution during the Euclidean time integration. Using Equation~(\ref{approx2}), we can calculate the Euclidean action and the result is
\begin{eqnarray}
S^{(1)}_{\mathrm{E}} = 4\pi^{2} \int_{0}^{X} \left( \rho^{3} V - \frac{3}{8 \pi} \rho \right) d \eta^{\mathfrak{Re}} \simeq - \frac{3}{8 m^{2} (\phi^{\mathfrak{Re}}_{0})^{2}} \sim - \frac{3}{16 V(\phi)}.
\end{eqnarray}
Therefore, as the vacuum energy becomes smaller and smaller, the probability gets larger and larger. This qualitative result is confirmed in more detailed calculations by Hartle, Hawking and Hertog \cite{Hartle:2007gi}\cite{Hartle:2008ng}.

\subsection{\label{sec:sym}Symmetric tachyonic potential}
To discuss potentials with local maxima, we now turn to the inverted square potential $V(\phi) = V_{0} - (1/2) m^{2} \phi^{2}$ as a model case. While analytic results are not available, we can still take a cue from the quadratic potential treated before, and form the hypothesis that
\begin{equation}
S_\text{E}^\mathfrak{Re} \simeq - \frac{3}{16} \frac{1}{V(\phi_0)}.
\label{eq:hyp}
\end{equation}
For the discussion of our numerical results, it is convenient to consider $S_\text{E}$ as a function of $\mu\phi_0$, where $\mu^{2}=m^{2}/V_0$, since according to \eqref{eq:hyp},
\begin{equation}
V_0 S_\text{E}^\mathfrak{Re} \simeq - \frac{3}{16} \frac{1}{1-\frac{1}{2} \mu^2\phi_0^2 }
\label{eq:hyp2}
\end{equation}
and hence independent of the parameters of the potential.

\begin{figure}
\begin{center}
\includegraphics[scale=1]{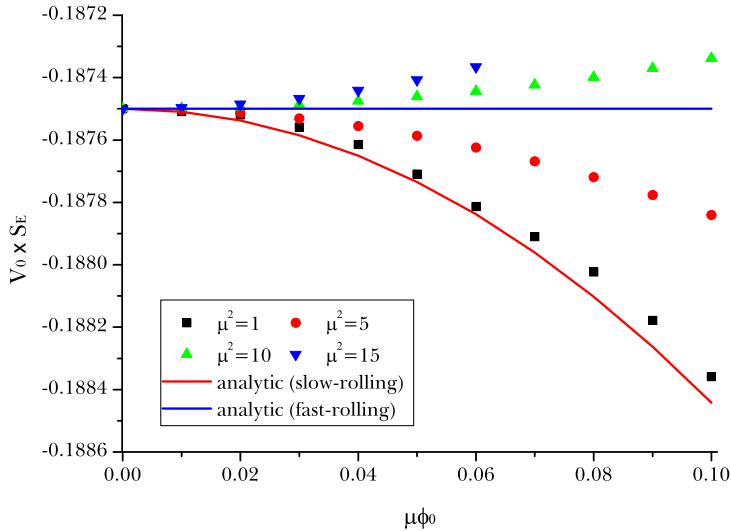}
\caption{\label{fig:transition2}Euclidean action $V_{0} S_{\mathrm{E}}$ as a function of $\mu\phi_{0}$. We searched for classicalized solutions for: $\mu^{2}=1, 5, 10, 15$. The red curve is the fitting function Equation~\eqref{eq:hyp2} as a function of $\mu \phi_{0}$ and the blue line is given by Equation~\eqref{eq:hyp3}.}
\end{center}
\end{figure}

\begin{figure}
\begin{center}
\includegraphics[scale=0.5]{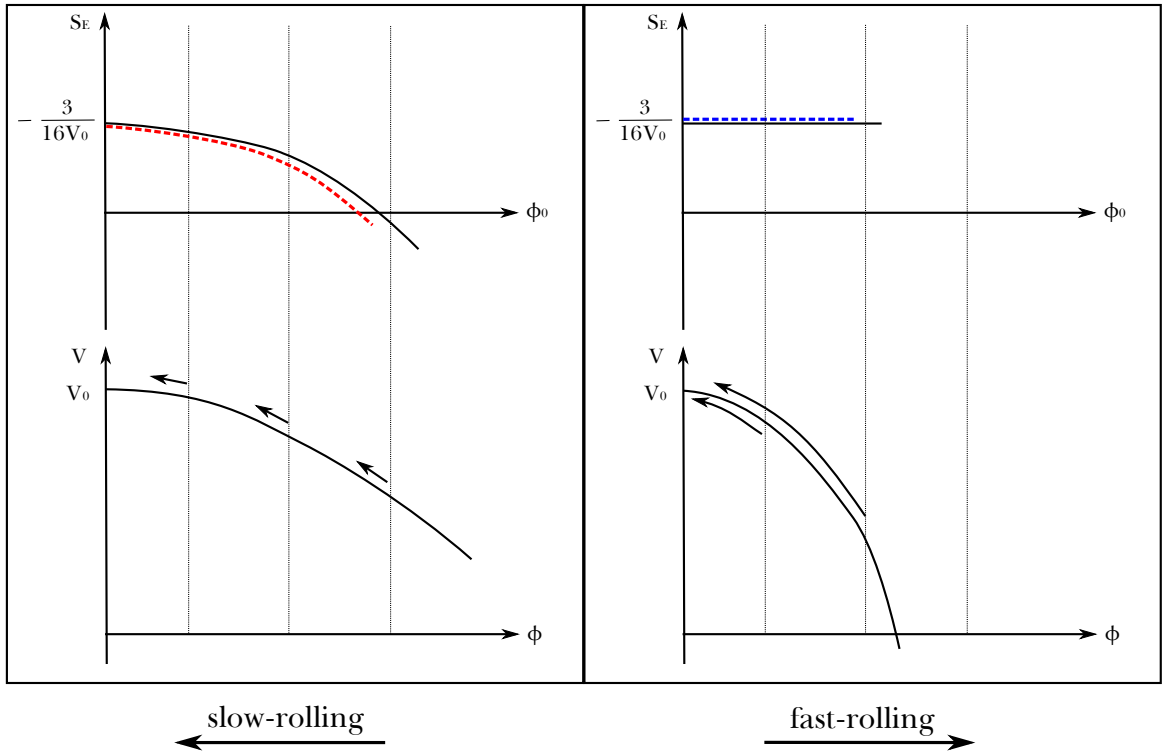}
\caption{\label{fig:conjecture4}Left: During Euclidean time, fields cannot roll up because the potential is sufficiently flat. Therefore, the final probability depends on the initial field values. Right: If the potential is sufficiently steep, then classical histories are allowed only for restricted field spaces. These initial field values roll up to the local maximum before classicalization, and hence the probability does not depend on the initial field values. (Dotted curves are fitting functions: Equations~\eqref{eq:hyp2} and \eqref{eq:hyp3}.)}
\end{center}
\end{figure}

Our numerical investigations show several things: First of all we do find instantons satisfying the classicality conditions for a wide range of parameters for the potential. Moreover, \eqref{eq:hyp2} describes the situation very well, as long as the potential is not too steep and the field evolves slowly as a consequence, see Figure~\ref{fig:transition2}. Deviations become apparent, however, as as $\mu^{2}$ increases,
and the real part of the action eventually becomes approximately independent of the starting modulus $\phi_0$ of the scalar field, see Figure~\ref{fig:transition2}.
In these instantons, the scalar field evolves quickly, and the Euclidean action is more accurately described by
\begin{equation}
V_0 S_\text{E}^\mathfrak{Re} \simeq - \frac{3}{16} \frac{V_0}{\max{V}}=- \frac{3}{16}.
\label{eq:hyp3}
\end{equation}
Instantons with this kind of behavior were also found in our previous work \cite{Hwang:2011mp} in the context of scalar tensor gravity, and were important to reach some of the conclusions of that work. The reason for the different behavior is that the slow-rolling instantons classicalize before ever reaching the maximum of the potential under evolution. The fast-rolling instantons reach the maximum and oscillate there before Lorentzian evolution sets in, therefore the maximum value of the potential is relevant for them, see Figure~\ref{fig:conjecture4}.

\begin{figure}
\begin{center}
\includegraphics[scale=1]{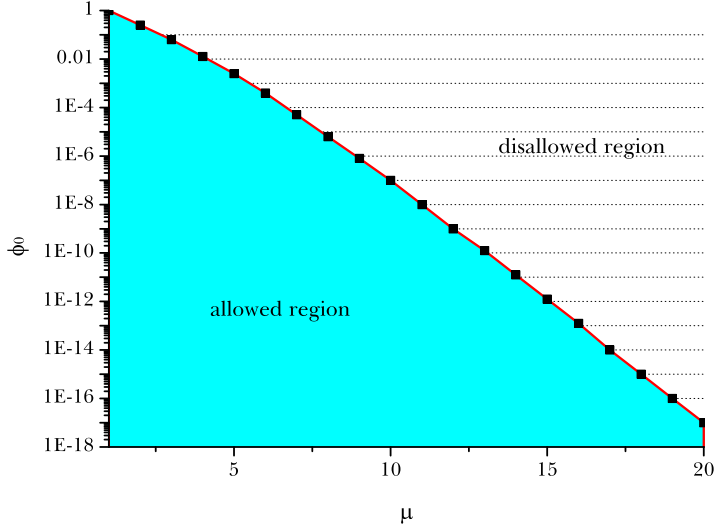}
\caption{\label{fig:HSY} Field space $\phi_{0}$ in which the classicalizing instantons are allowed, as a function of $\mu$: As $\mu$ increases, the allowed field space exponentially decreases.}
\end{center}
\end{figure}

There is another surprising difference between the slow-rolling and the fast-rolling instantons, namely the scaling of the region in which we find instantons satisfying the classicality condition. For any $\mu$ we find a symmetric region around the maximum of the potential such that there are classical histories starting with $\phi_0$ in that region. For $\mu^{2}\lesssim 1$, there is no solution when $V(\phi_{0}) < 0$, and hence $\Delta \phi_{0}$ should be less than the order of $\mu^{-1}$. We find approximately
\begin{equation}
\Delta \phi_0 \propto \begin{cases}
\mu^{-1} & \text{for\;\;\;} \mu^{2} \lesssim 1\\
\exp{- c \mu}& \text{for\;\;\;} \mu^{2} \gtrsim 10
\end{cases}
\label{eq:deltaphi}
\end{equation}
with $c$ given by (see Figure~\ref{fig:HSY})
\begin{equation}
c\approx 0.916 \ln 10 \approx 2.11.
\end{equation}

Let us finish by discussing the numerical results in terms of probabilities. Of particular interest are the probabilities $P_{\,\text{R}}$, $P_{\,\text{L}}$ for obtaining a classical universe with a right-rolling, respectively left-rolling, scalar field. Since the system is invariant under $\phi\leftrightarrow -\phi$, these probabilities are certainly equal.
For the slow-rolling case, we have
\begin{align}
P_{\,\text{L/R}} &\simeq  \int_{0}^{\Delta \phi_{0}} \exp{ \left( \frac{3}{8 V_{0}} + \frac{3}{16 V_{0}} \mu^{2} \phi_{0}^{2} \right)} d\phi_{0}\\
&\simeq  \frac{1}{\mu}\exp\left(\frac{3}{8 V_{0}} \left( 1 + \epsilon \right) \right),
\end{align}
where $\epsilon$ is order of $\mu$, and hence small in the slow-rolling case.
For the fast-rolling case  ($\mu^{2} \gtrsim 10$), we have
\begin{align}
P_{\,\text{L/R}} &\simeq  \exp{\frac{3}{8 V_{0}}} \int_{0}^{\Delta \phi_{0}}  d\phi_{0}\\
&\simeq  \exp\left[ \frac{1}{V_0}\left(\frac{3}{8}-c m \sqrt{V_{0}}\right)\right].
\end{align}
If the mass $m$ is sufficiently large, the probability of classicalized instantons can be exponentially suppressed. This can be very important when comparing probabilities for different types of instantons when the potential has a more complicated shape.

\subsection{\label{sec:run}Run-away potential}

Let us first consider the scaling
\begin{eqnarray}
\eta = D \tilde{\eta}, \;\;\; \rho = D \tilde{\rho}, \;\;\; V = D^{-2} \tilde{V},
\end{eqnarray}
where $D$ is a non-zero constant. Then we can easily show that
\begin{equation}
S_\text{E} = D^{2} \tilde{S}_\text{E},
\end{equation}
and hence the equations of motion are invariant via the scaling.
Therefore, if there is a fuzzy instanton solution, then there is also a fuzzy instanton solution for the scaled system, and vice versa.

Keeping this in mind, for a given exponential type potential
\begin{eqnarray}\label{eq:exp}
V(\phi) = A e^{-C \phi},
\end{eqnarray}
there is/is not a fuzzy instanton solution at $\phi^\mathfrak{Re}(\eta=0) = s_{0}$, if and only if there is/is not a fuzzy instanton solution with the redefined scalar field $\psi \equiv \phi-s_{0}$ and the redefined potential
\begin{eqnarray}
\tilde{V}(\psi) = e^{-C \psi},
\end{eqnarray}
by choosing the scaling parameter $D^{-2} = A e^{-C s_{0}}$.
This shows that to find classical histories for the potential in
Equation~(\ref{eq:exp}), we can focus on the case $\phi^\mathfrak{Re}(\eta=0)=0$.

\begin{figure}
\begin{center}
\includegraphics[scale=1]{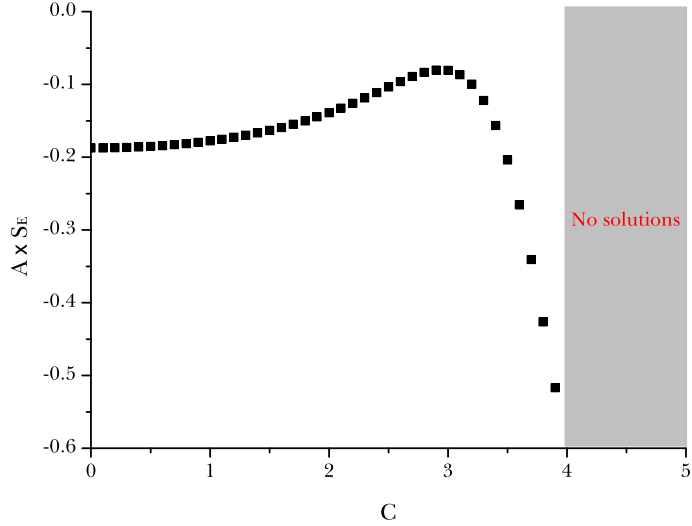}
\caption{\label{fig:exponential}Euclidean action for the exponential potential $V(\phi)=e^{-C\phi}$, around $\phi^{\mathfrak{Re}}(\eta=0)=0$. Around $C\cong3$, there appears qualitatively different behavior. If $C>4$, then classicalized solutions are not allowed.}
\end{center}
\end{figure}

We have searched for fuzzy instantons for this potential, using
numerics. By virtue of the above scaling argument, we can set
$\phi^\mathfrak{Re}(\eta=0)=0$ and then tune $X$ and $\phi^\mathfrak{Im}(\eta=0)$ to find
the classicalized solution. It turned out that very high numerical
accuracy was needed to reliably distinguish classical from
non-classical solutions. The result was very surprising. It is
depicted in Figure~\ref{fig:exponential}, where we have plotted the Euclidean action of
classical histories versus the parameter $C$ of the potential. There
is a critical value $C\simeq 4$ above which no classicalized
solutions exist. Due to the scaling argument, this result extends to
histories with arbitrary initial value for the field. This has
interesting implications for the stabilization probability of the HBSV
model discussed in the next section, and potentially for moduli
stabilization in general.

%In conclusion, for the potential in Equation~(\ref{eq:exp}), to search the existence of the fuzzy instanton, it is sufficient to see when the field becomes zero.

%For numerical searching, if $\phi_{0}=0$, then we tune $X$ and $\phi^{\mathfrak{Im}}(\eta=0)$ to find the classicalized solution. Figure~\ref{fig:exponential} is the result. For $C > 4$, there are no classicalized solutions.

Let us try to explain using analytic arguments. We can approximate around $\phi_{0}$ by the quadratic type potential
\begin{eqnarray}
\tilde{V}(\phi) = \frac{1}{2} \left( AC^{2}e^{-C\phi_{0}} \right) \left( \phi - \phi_{0} - \frac{1}{C} \right)^{2} + \frac{Ae^{-C\phi_{0}}}{2},
\end{eqnarray}
since $V(\phi_{0})=\tilde{V}(\phi_{0})$, $V'(\phi_{0})=\tilde{V}'(\phi_{0})$, and $V''(\phi_{0})=\tilde{V}''(\phi_{0})$. Hence, as long as $\phi - \phi_{0}$ is sufficiently smaller than $1$, then the approximated potential $\tilde{V}$ will give physically same results to those of $V$. If we consider the slow-rolling case, then such an approximation is sufficiently fine. Therefore, the potential now looks like
\begin{eqnarray}
\tilde{V}(\psi) = \frac{1}{2} m^{2} \psi^{2} + V_{0},
\end{eqnarray}
where $\psi = \phi - \phi_{0} - 1/C$.

Revisiting the results of Hartle, Hawking, and Hertog \cite{Hartle:2008ng}:
\begin{enumerate}
\item If $m^{2} > 6 \pi V_{0}$, or equivalently, $C^{2} > 3\pi$, then there is a cutoff $\psi_{c}$ such that if $|\psi|<\psi_{c}$, then there is no classicalized solution. Here, $\psi_{c} \sim \sqrt{3/4\pi} \times \mathcal{O}(1)$.
\item Otherwise, if $C^{2} \leq 3\pi$, then there is no cutoff $\psi_{c}$.
\end{enumerate}
Therefore, we can conclude that
\begin{itemize}
\item For $C > \sqrt{3\pi} \sim 3$ cases, if $\psi_{c} > 1/C$ or equivalently $C > \sqrt{4\pi/3} \times \mathcal{O}(1) \sim 2 \times \mathcal{O}(1)$, then there will be no classicalized solution.
\item If $C \leq \sqrt{3\pi}$, then there are always fuzzy instantons.
\end{itemize}
Therefore, when we compare the numerical results, the order one constant $\mathcal{O}(1)$ is approximately $2$. In addition, we can explain the behavior: around $C = \sqrt{3\pi} \cong 3$, there appears a qualitatively different phase.

%Equation~(\ref{eq:exp}) can be rescaled by $A \rightarrow 1$ and the analysis around $\phi_{0}$ can be shifted to $\phi_{0} \rightarrow 0$, with a proper field redefinition: there is a classicalized solution for Equation~(\ref{eq:exp}) around $\phi_{0}$ if and only if there is a classicalized solution for $e^{-C\phi}$ around $\phi_{0}=0$.

For a typical form of the moduli potential
\begin{eqnarray}
V(\phi) = \sum_{i} f_{i}(\phi) e^{-C_{i}\phi},
\end{eqnarray}
where $f_{i}$s are polynomials of $\phi$ and $C_{i}$s are constants, we require the condition
\begin{eqnarray}
\min{C_{i}} \gtrsim 4.
\end{eqnarray}
If this condition does not hold, then the no-boundary measure cannot explain the stabilization of moduli fields.

\section{No-boundary measure in string theory}

\subsection{Stabilization of moduli fields}

%\subsubsection{The HBSV model}

\begin{figure}
\begin{center}
\includegraphics[scale=0.55]{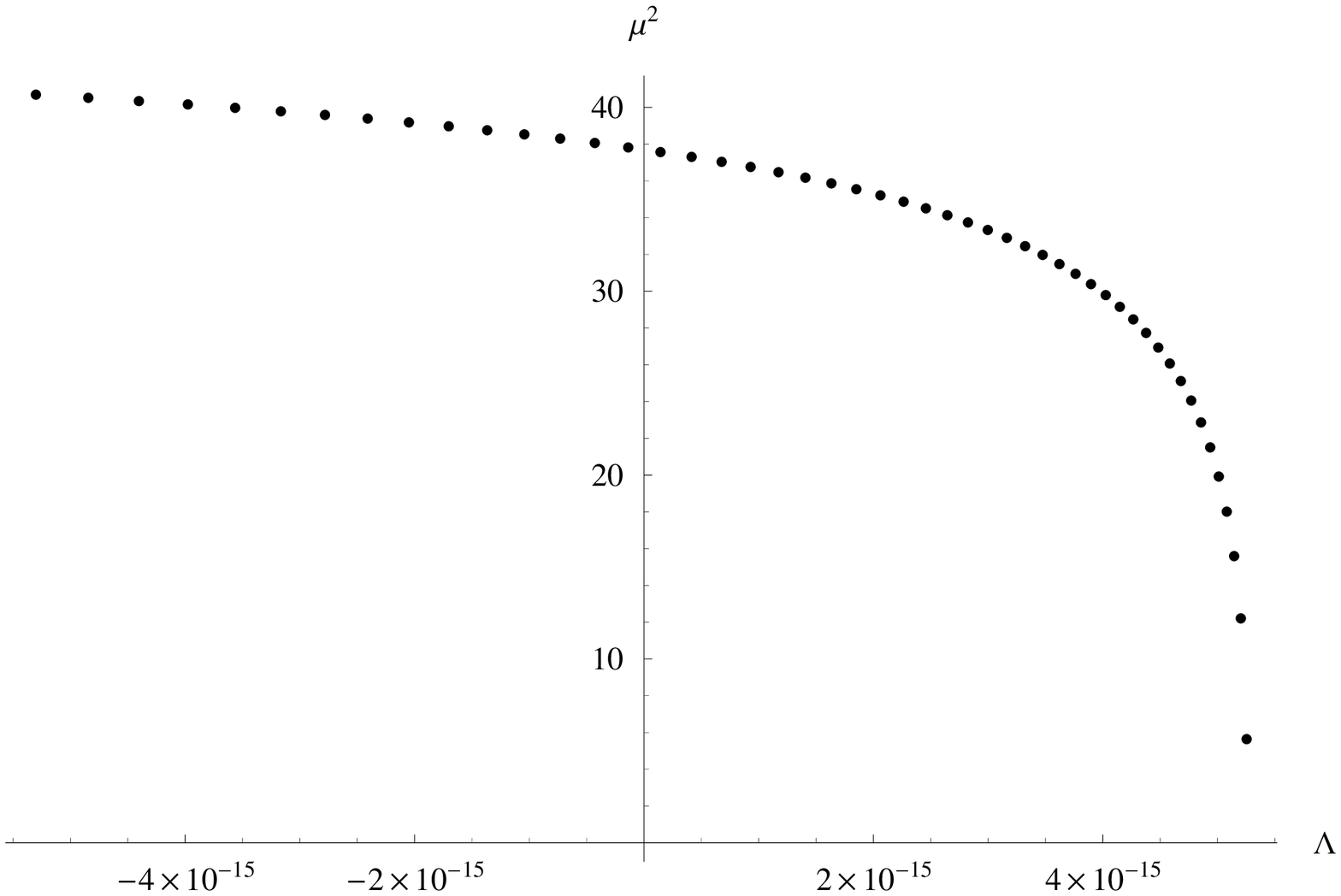}
\includegraphics[scale=0.7]{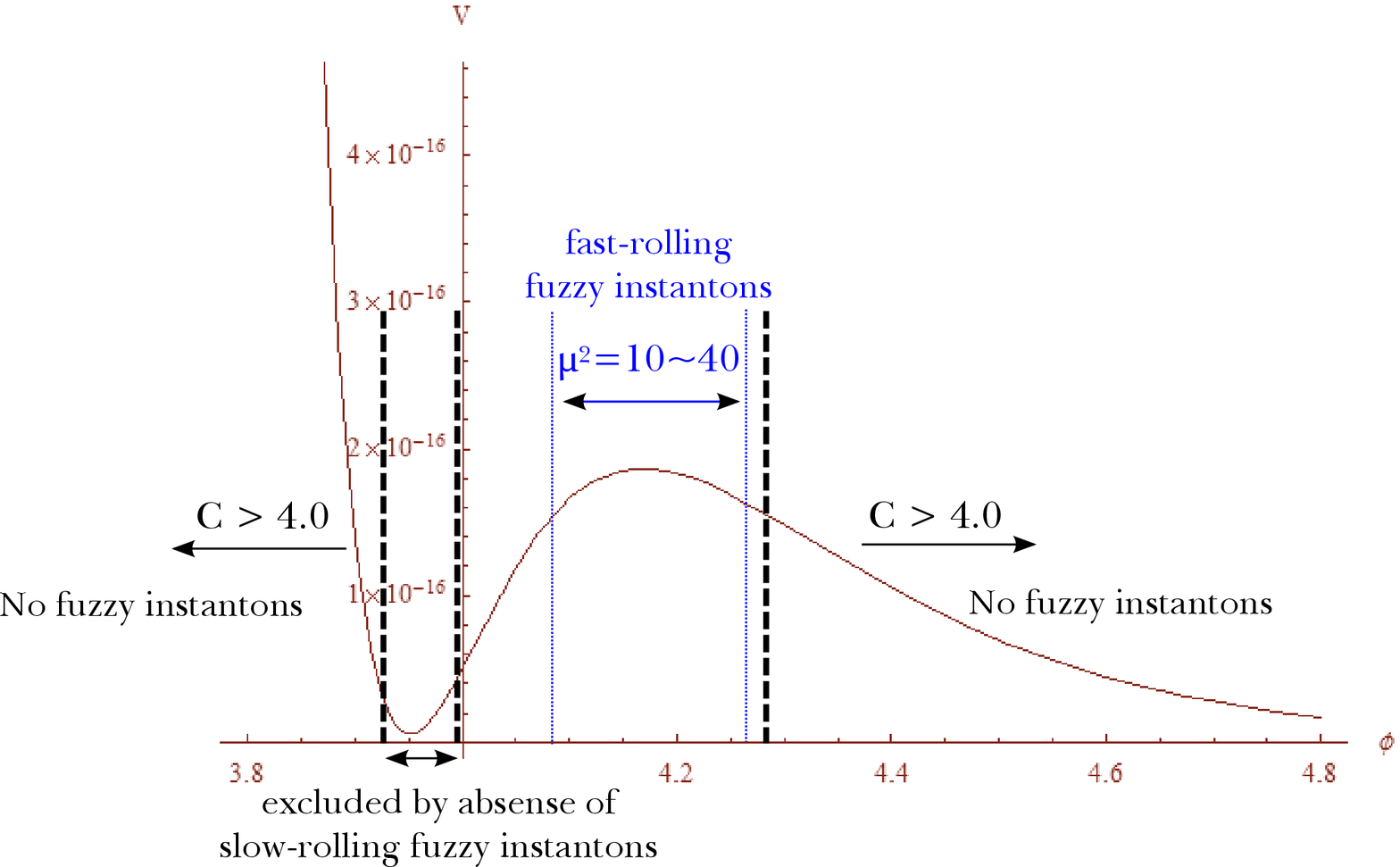}
\caption{\label{fig:mu_Lambda}An example of the HBSV model: $R=0.01$, $e=0.1$, $\Lambda^{(6)}=0.01$, $n=225$. Upper: We plotted $\mu^{2}$ at the local maximum as a function of $\Lambda = 8\pi V_{\mathrm{min}}$. Therefore, typical $\mu^{2}$ is greater than $10$ and hence the probability will be almost flat. Lower: A typical potential. Around the local maximum, $\mu^{2} \gtrsim 10$ such that fast-rolling instantons can classicalize there. Instantons cannot classicalize at large values of $\phi$ since the potential is $\sim \exp -C\phi$ and $C = \sqrt{8\pi} > 4.0$. Therefore run-away solutions cannot be generated in this way.}
\end{center}
\end{figure}

\begin{figure}
\begin{center}
\includegraphics[scale=0.7]{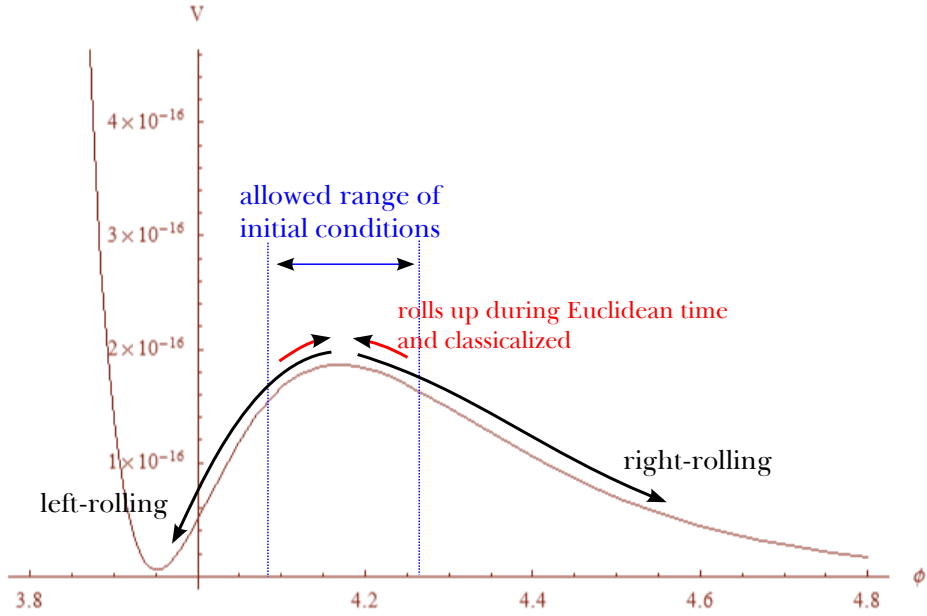}
\caption{\label{fig:6Dpotential2}Conceptual discussion on the moduli stabilization problem. As discussed in Figure~\ref{fig:mu_Lambda}, the initial conditions are allowed only by the blue colored region. Between the allowed region, a fuzzy instanton can start and role up around the local maximum (red arrows) along the Euclidean time. After the turning point, it can role to the left side or the right side (black arrows) along the Lorentzian time. Approximately, the probability is almost half, and hence it can explain the stabilization of the moduli field.}
\end{center}
\end{figure}

Let us analyze the HBSV model in detail (Figure~\ref{fig:mu_Lambda}).
\begin{enumerate}
\item For typical examples, $\mu^{2}$ at the local maximum is greater than $10$. Therefore, around the local maximum, in many cases, it follows almost flat probability distribution (fast-rolling fuzzy instantons).
\item For the run-away direction, it is approximately $e^{-\sqrt{8\pi} \phi}$. However, in our model, $\sqrt{8\pi} > 4.0$. Therefore, there are no classical solutions for the run-away direction.
\item Around the local minimum, there is no classical solution (according to Hartle, Hawking, and Hertog). Also, there is no classical solution for the left side of the local minimum, since $\sim e^{-3\sqrt{8\pi}\phi}$.
\end{enumerate}
Therefore, for our specific model, we can explain the stabilization of the moduli field. We denote the conceptual picture in Figure~\ref{fig:6Dpotential2}. As we discussed, the initial conditions are allowed only around the local maximum. Around the local maximum, a fuzzy instanton can start and role up near the local maximum (red arrows) along the Euclidean time. After the turning point, it can role to the left side or the right side (black arrows) along the Lorentzian time. Approximately, the probability of each side is almost half, and hence it can explain the stabilization of the moduli field.

Perhaps, this can be generalized.
\begin{itemize}
\item Small $\mu^{2}$ is related to small slow-roll parameters. Therefore, slow-rolling fuzzy instantons needs some fine-tuning for the potential.
\item For run-away potential $e^{-C \phi}$, $C$ will be on the order of Planck units ($\sim \sqrt{8\pi}$). Therefore, in fact, the condition $C > 4.0$ is quite general for moduli stabilization models, since the typical dimensions of such a theory should be Planck units.
\item However, if we have to consider many numbers of moduli fields, then the stabilization of \textit{all} moduli fields will be a difficult problem. If there are $N$ `coherent' fields with potential $V = \sum_{i} e^{-C\phi_{i}}$, then it will be approximated by the potential $N e^{-C/\sqrt{N} \psi}$ with a single field $\psi=\phi_{i}\sqrt{N}$. Hence, even though $C>4$, for large $N$, $C/\sqrt{N}$ can be sufficiently small. Therefore, the stabilization is highly non-trivial. We will leave this for future work.
\end{itemize}

\subsection{Bayesian inference: \textit{A priori} probability of cosmological constant}

Up to now we have studied the probability distribution for the motions of the moduli field. Now we would like to discuss the flux, since it determines, among other things, the cosmological constant. In the HBSV model, the flux is simply an input. But in view of applications in the context of cosmic landscape, there may be a continuous function of a field that varies the flux. Of course, we do not know how to construct the exact landscape potential and it highly depends on the underling theory. In this context, it may be useful to regard it as a random variable. Alternatively, we can still consider it fixed but unknown, and consider our belief in its various values in the sense of Bayes.

In either case, we have
\begin{eqnarray}
P(n|h) = \frac{P(h|n)P(n)}{P(h)}
\end{eqnarray}
where $n$ denotes the value of charge in the HBSV model, and $h$ is a (stabilized) history, or a family of such histories. What is given to us by the no-boundary proposal is $P(h|n)$.
Note that the normalization of the probability becomes
\begin{eqnarray}
\sum_{n}\left(\sum_{\mathrm{h:stable}} P(h|n) + \sum_{\mathrm{h:unstable}} P(h|n)\right) = 1
\end{eqnarray}
and from the discussion of the previous section, we obtain
\begin{eqnarray}
\sum_{\mathrm{h:stable}} P(h|n) \simeq \sum_{\mathrm{h:unstable}} P(h|n).
\end{eqnarray}
However, in general $P(\text{stab}|n) \neq P(\text{stab}|n')$ for $n\neq n'$ and it depends on the Euclidean action.

Let us assume we observe stable moduli. Based on this observation, we obtain a probability distribution for $n$,
\begin{equation}
P(n|\text{stab})=\sum_{h: \text{stable}} \frac{P(h|n)P(n)}{P(h)}
%\label{eq:}
\end{equation}
$P(h)$ is approximately constant on the set of stabilized histories, and we also assume $P(n)$ is constant in the absence of any better idea. Then
\begin{equation}
P(n|\text{stab})\propto \sum_{h: \text{stable}} P(h|n)
%\label{eq:}
\end{equation}
We can calculate the probability density $p(h_{\phi_{0}}|n)$, where $h_{\phi_{0}}$ is a classical history labeled by the initial condition $\phi_{0}$. It is approximately given by the intergrand in \eqref{eq:prob}.
Then we can define
\begin{equation}
P(\text{stab}|n) \equiv \int_{C \rightarrow \mathrm{L}} p(h_{\phi_{0}}|n) d\phi_{0}.
\end{equation}
where $C$ is the set of initial conditions with classical histories.
Then from our previous discussion
\begin{equation}
P(n|\text{stab}) \propto  P(\text{stab}|n).
\end{equation}
For each different $n$, each stabilized history will have a specified potential energy.  Therefore, we also obtain a distribution $P(\Lambda|\text{stab})$, where $\Lambda$ is the effective cosmological constant. Therefore, this will give the \textit{a priori} probability of the cosmological constant.

To approximately evaluate the probability distribution, we will use the relation:
\begin{eqnarray}
P_{\,\mathrm{L}} \simeq \exp \left( \frac{3}{8 V_{\max}} \left( 1 + \epsilon \right) \right) \Delta \phi_{0}
\end{eqnarray}
where $\epsilon$ may be non-zero for instantons that are slow-rolling. To see the qualitative properties, it is not unreasonable to ignore $\epsilon \ll 1$. We have to consider $\Delta \phi_{0}$. However, if it is not measure zero, then it will not have a significant effect. Therefore, the crucial point is the relation between the local minimum and the local maximum of the potential, where it crucially depends on the details of the potential.

\begin{figure}
\begin{center}
\includegraphics[scale=0.55]{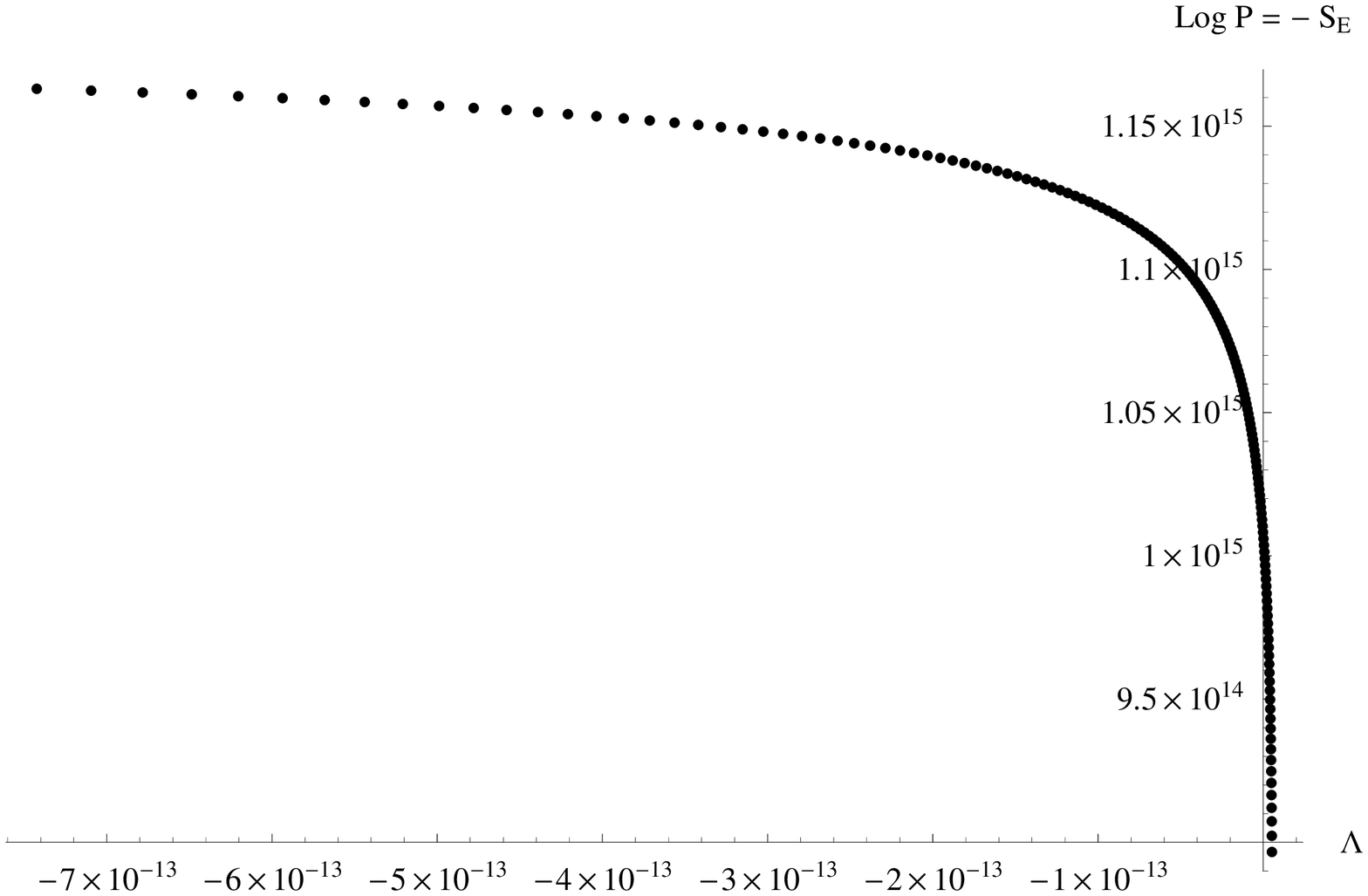}
\includegraphics[scale=0.55]{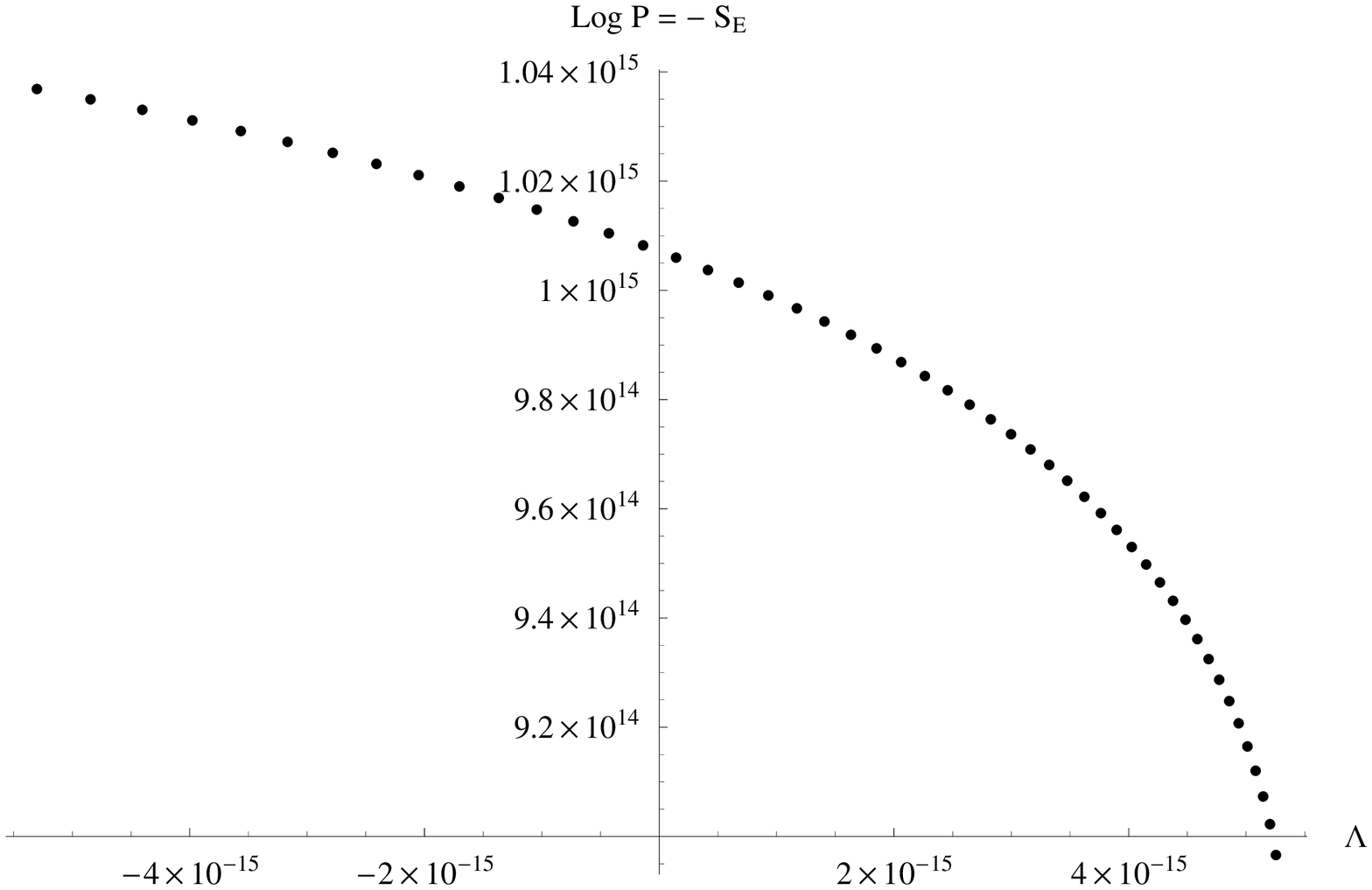}
\caption{\label{fig:landscape}$\log P(\Lambda|\text{stab}) \simeq - S_{\mathrm{E}} \simeq 3/16V(\phi_{\mathrm{M}})$ as a function of $\Lambda = 8 \pi V(\phi_{\mathrm{m}})$. We used the condition $R=0.01$, $e=0.1$, and $\Lambda^{(6)}=0.01$. The former plot is the overall distribution for $100 < n < 259$. The latter plot is the distribution near $\Lambda=0$ $(210 < n < 259)$.}
\end{center}
\end{figure}

Figure~\ref{fig:landscape} shows the typical distribution on $\log P(\Lambda|\text{stab})$ for an HBSV model. In the model, we used the condition $R=0.01$, $e=0.1$, $\Lambda^{(6)}=0.01$. Then the allowed $n$ is less than $259$. The former plot is the overall distribution for $100 < n < 259$. The latter plot is the distribution near $\Lambda=0$ $(210 < n < 259)$.

We can obtain two general conclusions:
\begin{enumerate}
\item Anti de Sitter vacua are preferred to de Sitter vacua.
\item However, if we restrict to $\Lambda \simeq 0$ so that the cosmological constant should be sufficiently small for anthropic requirements, then we will see effectively flat a priori probability.
\end{enumerate}
Therefore, $\Lambda=0$ is no longer a special point in terms of Euclidean quantum gravity.

\subsection{Generalization to cosmic landscape: AdS catastrophe?}

\paragraph{Dead de Sitter catastrophe of multiverse: Susskind and Page's arguments}

First, let us summarize the assumptions and results of Dyson, Kleban, and Susskind \cite{Dyson:2002pf}:
\begin{description}
\item[Assumption~$1$.] The time evolution of multiverse is \textit{unitary} for an observer's causal patch.
\item[Assumption~$2$.] There is a fundamental cosmological constant larger than $0$.
\end{description}
According to Assumption~$1$, the probability of the vacuum energy of an observer in the multiverse (or manyworld \cite{Nomura:2011dt}) will proportional to the exponential of the entropy: $P \sim \exp-2S_{\mathrm{E}}$. According to Assumption~$2$, the most probable cosmological constant is the smallest vacuum energy $\Lambda > 0$. Then, in the full phase space, the probability that one will see inflation with the vacuum energy $V_{0}$ is approximately
\begin{eqnarray}
P \simeq \exp{\left( \frac{1}{V_{0}} - \frac{1}{\Lambda} \right)}.
\end{eqnarray}
For our universe, $V_{0} \simeq 10^{-10}$ and $\Lambda \simeq 10^{-120}$. If we consider $10^{500}$ numbers of vacua, then $\Lambda \simeq 10^{-500}$ is not a strange number. Then, $P \simeq 0$. According to Dyson, Kleban, and Susskind \cite{Dyson:2002pf}, therefore, the most dominant position in the phase space of the multiverse is the dead de Sitter. According to Page \cite{Page:2006hr}, the dead de Sitter is filled by not human observers but freak observers, for example, so-called Boltzmann Brains. To solve this problem, we have to include anti de Sitter vacua and a proper measure should be defined so that de Sitter vacua should decay to anti de Sitter vacua before Boltzmann brains are dominant.

Therefore, introduction of anti de Sitter region (dropping the Assumption~$2$) resolves such a problem. Now what we want to discuss is that the introduction of anti de Sitter vacua can generate a difficult problem for the no-boundary measure.

%This is due to the property of entropy of de Sitter spaces and the property of Euclidean quantum gravity. Multiverse is constructed by instantons (pocket universes) tunneling from something (inflating background). On the other hand, the no-boundary measure explains instantons (fuzzy instantons) from nothing (no-boundary). The former uses Lorentzian time, while the latter uses from Euclidean time to Lorentzian time. So, for the former, the local minimum is important to calculate entropy, while the local maximum is important for the latter. Therefore, the dead de Sitter catastrophe is changed to the anti de Sitter catastrophe, if the anti de Sitter vacuum is the nearest vacuum of the smallest local maximum with positive vacuum energy.

\paragraph{Anti de Sitter catastrophe and the no-boundary measure}

If there is a local maximum with a very small vacuum energy of a potential that is allowed from string theory, then around the point will give $\mu^{2} = m^{2}/V_{0}$ parameter. As we discussed, unless $m$ is super-Planckian, the probability will be approximately $\sim \exp 1/V_{0}$.

\begin{figure}
\begin{center}
\includegraphics[scale=0.7]{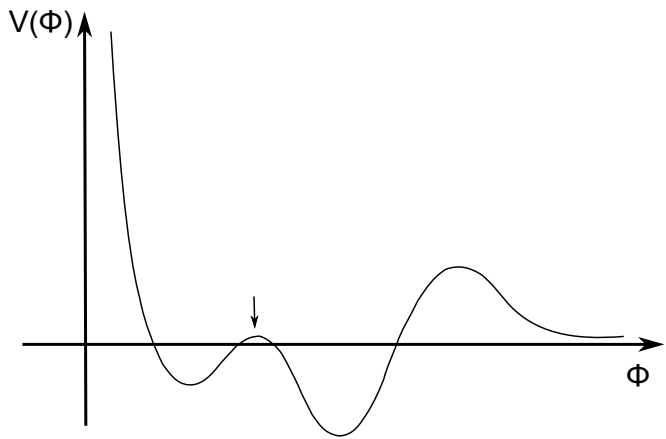}
\caption{\label{fig:landscape_example}An example of a potential that can be allowed from cosmic landscape. The local maximum near zero vacuum energy (arrow) is the most probable region, and the nearest vacua of the local maximum will probably be anti de Sitter.}
\end{center}
\end{figure}

The question is how small the vacuum energy will be. If the cosmic landscape allows $10^{500}$ vacua, then the possible local maxima can be on the order of $V_{0} \sim 10^{-500}$. Then, the position $V_{0} \sim 10^{-500}$ is the most favorable position in terms of the no-boundary measure. Then, after the turning time, the scalar field will roll to the nearest vacua of the local maximum. If $V_{0}$ is very close to zero, then the nearest vacuum will probably be anti de Sitter (Figure~\ref{fig:landscape_example}). Therefore, it seems that the natural version of string theory seems to prefer not de Sitter but anti de Sitter vacua, on the order of $\exp 10^{500}$. We will call this \textit{anti de Sitter catastrophe}.

We suggest possible resolutions of the anti de Sitter catastrophe.
\begin{description}
\item[Hypothesis~$1$.] There can be some fundamental limitations on the potential. For example, there can be a fundamental principle that the mass of the local maximum should be super-Planckian. Also, there can be a limitation that whenever the local maximum has sufficiently small vacuum energy, the nearest vacuum should be de Sitter.
\end{description}
These possibilities can resolve the anti de Sitter catastrophe; however, we think that these resolutions are quite \textit{ad hoc}.

Rather, we suggest two probable resolutions.
\begin{description}
\item[Hypothesis~$2$.] The bottom-up probability prefers anti de Sitter vacua to de Sitter vacua. However, in terms of \textit{anthropic} considerations, a universe should experience inflation. However, such anti de Sitter fuzzy instantons cannot experience sufficient e-foldings. Then, the universe will not have anthropically viable structures and cannot include human-like observers.
\end{description}
This possibility can be falsifiable. We have to check whether such an anti de Sitter universe can include Boltzmann-brain--like freak observers or not. If it is possible, then we can compare the number of freak observers of the anti de Sitter universe and the number of human-like observers in our universe. After we weight the bottom-up probability, if the former is dominant over the latter, then Hypothesis~$2$ cannot be true.
\begin{description}
\item[Hypothesis~$3$.] Even though the bottom-up probability prefers an anti de Sitter vacuum, there is a small probability that a universe can experience eternal inflation. Then, although it is a small probability, after a long time, the volume of the eternally inflating histories will be dominant over the other anti de Sitter histories, and the anti de Sitter histories will experience a big crunch in a finite time. Then, eventually, the eternally inflating universe will remain. Then, it can generate de Sitter vacua as pocket universes.
\end{description}
Then, the distribution of the pocket universes will follow the measure of eternally inflating multiverse. Now it is not clear whether there will remain a signature of the no-boundary measure through the eternal inflation, or it is entirely erased.
%This implies that the initial information on the no-boundary measure will eventually be erased via eternal inflation.

In conclusion, we obtained a result that the no-boundary bottom-up measure prefers not de Sitter but anti de Sitter vacua exponentially. One way to resolve this is the anthropic argument and the other way is to introduce eternal inflation.
%Once an observer can be viable in the nearest anti de Sitter vacuum, then perhaps the typicality of the observer can be highly emphasized via the exponential factor. Therefore, the latter possibility seems to be more natural. However, the latter idea has another theoretical problem for the multiverse. For example, the measure of the multiverse is not well defined and other assumptions are requested.
For this issue, we may not be able to say any concrete things, since our knowledge for all over the landscape is absent. Rather, our cautious conclusion is that the no-boundary measure potentially have a problem of anti de Sitter catastrophe, and this is another version of the dead de Sitter catastrophe of multiverse.
% our cautious conclusion is that if one requests a natural explanation for our universe, then \textit{we have to encounter eternal inflation and multiverse, even though we begin from the no-boundary wave function of the universe}. However, if we accept some chances to explain our universe, the no-boundary measure is still a well-defined idea.

\section{Discussion}

In this paper, we investigate the no-boundary measure in string theory: in the context of moduli stabilization, flux compactification, and cosmic landscape. The simple model that compactifies $6$D to $4$D (HBSV model) is used as a concrete toy model.

\begin{description}
\item[Section~\ref{sec:sym}:] To study fuzzy instantons, the study of the tachyonic potential around the local maximum is very important. If the ratio between the negative mass square $m^{2}$ and the vacuum energy $V_{0}$ decreases, then we observe slowly-rolling fuzzy instantons. In this case, the probability depends on the initial conditions. We call this case \textit{slow-rolling fuzzy instantons}. On the other hand, if the ratio sufficiently increases, the field relatively quickly moves and hence the dependence on the initial conditions disappears. We call this case \textit{fast-rolling fuzzy instantons}.
\item[Section~\ref{sec:run}:] For an exponential type run-away potential $e^{-C \phi}$, if the coefficient $C$ is larger than $\sim 4$, then there is no run-away classicalized solution for the no-boundary measure.
\end{description}

These two observations are applied to the HBSV model: classical histories are allowed only for the local maximum, and the probability will not crucially depend on the initial conditions (fast-rolling fuzzy instantons). Therefore, \textit{the no-boundary measure can explain the stabilization of the HBSV model}. Moreover, we can assert the \textit{a priori} probability of the cosmological constant. It naturally prefers to be anti de Sitter, but a zero cosmological constant is not a special or singular region.

Finally, we try to generalize to the context of the cosmic landscape. Perhaps, the no-boundary measure extremely prefers anti de Sitter to de Sitter. This can be a fundamental property of the no-boundary measure. If we believe the principle that \textit{a probability of a given gravitational system is determined by the entropy, which is the same as the Euclidean action}, then it will be a natural consequence.

However, this crucially depends on the details of the potential and the choice of the boundary condition of the universe. Thus, we are not ready to conclude with definite results, or to say for certain whether the no-boundary measure is consistent with string theory or not. We have already obtained some evidence that the no-boundary measure partly explains the stabilization of the moduli fields and the dilaton field \cite{Hwang:2011mp}. Of course, we have to generalize not only for the single field case, but also for multi field cases. We need further study for these issues.

\section*{Acknowledgment}

The authors would like to thank Ewan Stewart for discussions and encouragement. DY and BHL were supported by the National Research Foundation of Korea(NRF) grant funded by the Korea government(MEST) through the Center for Quantum Spacetime(CQUeST) of Sogang University with grant number 2005-0049409. DH was supported by Korea Research Foundation grants (KRF-313-2007-C00164, KRF-341-2007-C00010) funded by the Korean government (MOEHRD) and BK21. HS was partially supported by the Spanish MICINN project No. FIS2008-06078-C03-03.

\end{document}